\documentclass{elsart}

\usepackage{amssymb}

\begin{document}

\begin{frontmatter}


 \title{Curci-Ferrari mass and the Neuberger problem}
 \author{A.C. Kalloniatis$^{a}$, 
L. von Smekal$^{b}$, 
A.G. Williams$^{a}$}
\thanks{Address for L. von Smekal after March, 2005: CSSM,
University of Adelaide, South Australia 5000, Australia.}
 \address{$^a$Centre for the Subatomic Structure of Matter,
University of Adelaide, South Australia 5000, Australia \\
$^b$Institute for Theoretical Physics III, University of
Erlangen-N\"urnberg, Staudtstrasse 7, Germany.}


\begin{abstract}
We study the massive Curci-Ferrari model as a starting point for defining 
BRST quantisation for Yang-Mills theory on the lattice.
In particular, we elucidate this proposal in light of
topological approaches to gauge-fixing and study   
the case of a simple one-link Abelian model. 
\end{abstract}

\thanks{Preprint numbers: ADP-04-26/T608; FAU-TP3-04/6.}
\end{frontmatter}


BRST symmetry has proven an invaluable tool in the 
perturbative quantisation of gauge theories \cite{Becchi:1975} 
so that its elevation to the non-perturbative level is clearly desirable.
For one, BRST methods
are crucial in the formulation of Schwinger-Dyson equations (SDEs)
in covariant gauges, which seem genuinely non-perturbative,
have been subject to study using various truncations
at this level\cite{Alkofer:2000}, and whose results are now subject to
comparisons with corresponding computations in Landau gauge
from lattice gauge theory (see, for example, \cite{Bowman:2004}).
However, it has been demonstrated that  
a standard formulation of BRST symmetry invoked for lattice fields rigorously
forces the partition function and path integrals
of BRST invariant operators with a BRST invariant measure
to vanish identically \cite{Neuberger:1986}. Rather than defining the 
configuration space in terms of some subset
with no Gribov copies \cite{Gribov:1977,Dell'Antonio:1991}, one sums
them all with alternating sign of the
Faddeev-Popov determinant (only for small fields about $A_{\mu}=0$
is this positive and thus the Jacobian of a change of variables)
and a complete cancellation takes place
giving the nonsensical result $0/0$ for lattice expectation values. 
This corresponds to attempting to resolve unity in the
Faddeev-Popov trick via the partition function of
a topological quantum field theory (TQFT) whose fields are
the $SU(n)$ group elements in the background of an external
gauge or link field \cite{Baulieu:1996}. 
Such integrals of BRST invariant observables are topological invariants,
the Euler character for $SU(n)$ for the partition function, which all
vanish giving the zeroes of the Neuberger problem.
Equivariant gauge-fixing \cite{Schaden:1998,Golterman:2004}
evades this no-go theorem through a sequential gauge-fixing via
coset space decomposition of $SU(n)$,
as in maximal Abelian gauge, so that
the submanifolds of $SU(n)$ implicit in this decomposition have
non-zero Euler character.
This formulation is however distant 
from covariant gauge Schwinger-Dyson equations. Significantly though, there are
quartic ghost couplings signalling that the damping term
in the action for scalar Nakanishi-Lautrup auxiliary fields $B$ is 
not BRST exact, $B^2 \neq s(\rm{something})$.
This also signals the break down of the Neuberger argument.

Quartic ghost couplings also arise in
generalisations of the BRST and anti-BRST symmetry of
Landau gauge through the so-called Curci-Ferrari (CF) ``gauges''
\cite{Curci:1976,Baulieu:1981,Thierry-Mieg:1985,Alkofer03}. They also allow for a massive 
vector field while retaining BRST- (though not gauge-)invariance. 
But nilpotency of the (anti-)BRST 
algebra and thus unitarity are lost to be recovered, along with
the original YM theory, in the massless limit. Nonetheless,
this limit gives a theory which is local, covariant, BRST invariant,
perturbatively renormalisable 
and close in spirit to Landau gauge.
We will elucidate these details in light of the topological approach
to gauge-fixing and the Neuberger problem in the following,
showing how these ``bugs'' possibly become features
which may enable a non-perturbative definition of BRST.

We assume antihermitian $SU(n)$ generators $T^a$
so that ghost components $C^a$ are hermitian
in order that $C(x)=C^a(x) T^a=-C^{\dagger}$. On the lattice ghosts
live on lattice sites $C_i$. In the continuum we deal
with $SU(n)$ gauge fields $A_{\mu}(x)=A_{\mu}^a(x) T^a$
while on the lattice we denote the link fields as
$U_{ij}$ for the link from lattice site $i$ to $j$,
$U_{ij}=P{\rm{exp}}(\int_{x_i}^{x_j} dz \cdot A)$.
The covariant derivative in the adjoint representation
is $D_{\mu} \cdot = \partial_{\mu}\cdot + [A_{\mu},\cdot]$.
We also require a Nakanishi-Lautrup auxiliary
$SU(n)$-algebra valued hermitian field $B(x)$ which lives on sites for
the lattice theory, $B_i$. Note that the antihermiticity
of the generators mean that the components of $B$ are imaginary.
Commutators and anti-commutators of variables will
be separately indicated by $[\cdot , \cdot]$ and 
$\{\cdot , \cdot \}$.

For the continuum theory we have the BRST and anti-BRST algebras 
\begin{eqnarray}
s A_{\mu} &=& D_{\mu}C, \ {\bar s} A_{\mu} = D_{\mu} {\bar C}, 
\label{contBRST} \\
s C &=& C^2 =\frac{1}{2}\{C,C\}, \ {\bar s} {\bar C} = {\bar C}^2 
=\frac{1}{2}\{{\bar C},\bar{C}\}  \\
s {\bar C} &=& B + \frac{1}{2} \{\bar{C},C\} , \ 
{\bar s}  C = - B + \frac{1}{2} \{C,\bar{C}\}.
\end{eqnarray}
where for the lattice Eqs.(\ref{contBRST}) are replaced by
\begin{equation}
sU_{ij}= C_i U_{ij} - U_{ij} C_j, \ \ {\bar s}U_{ij} 
= {\bar C}_i U_{ij} - U_{ij} {\bar C}_j \label{barsAU}
.
\end{equation}
With these, the gauge-fixing is symmetric under 
$C\rightarrow {\bar C}, \ {\bar C}\rightarrow -C $. 
We have then
$s {\bar C} + {\bar s} C - \{\bar{C},C\}=0$
which can be geometrically interpreted as the vanishing of a curvature 
in the extended space and so both ghosts and anti-ghosts
are Maurer-Cartan one-forms.
We still have some freedom in specifying the
variations of the auxiliary field $B$.
We choose
\begin{eqnarray}
sB &=& m^2 C -\frac{1}{2} [B,C] + \frac{1}{8} [\bar{C},\{C,C\}] \nonumber\\
{\bar s}B &=& m^2 {\bar C} -\frac{1}{2} [B,{\bar C}] 
- \frac{1}{8} [C,\{{\bar C},\bar{C}\}] .
\end{eqnarray}
An $Sp(2)$ group with generators $\sigma^+,\sigma^0,\sigma^-$
specified by $\sigma^i A_{\mu}=\sigma^i B=0$ and 
\begin{eqnarray}
\sigma^- C = {\bar C}, \sigma^0 C = C, \sigma^+ C=0, \
\sigma^- {\bar C} = 0, \sigma^0 {\bar C} = -{\bar C}, \sigma^+ {\bar C}=C
\end{eqnarray}
enables us to replace nilpotency by the relations:
\begin{equation}
s^2 = m^2 \sigma^+ , {\bar s}^2 = -m^2 \sigma^-  ,
s{\bar s} + {\bar s}s = -m^2 \sigma^0 .
\end{equation}
The gauge-fixing of the Yang-Mills action can be
achieved by the addition of a (anti-)BRST-invariant but not BRST-exact
action, namely
\begin{equation}
S_{GF}=(s {\bar s} - m^2)W, 
\label{Sgf}
\end{equation}
where for the continuum (respectively lattice) theory 
the most general choice for $W$ is
\begin{equation}
W_{\rm{cont}} =\frac{1}{2}{\rm{Tr}}\int d^4x [(A_{\mu})^2 - \xi {\bar C} C], \
W_{\rm{lat}}=\frac{1}{2} \sum_{ij} {\rm{Tr}}[ {\rm{Re}} U_{ij} 
- \xi {\bar C}_i C_i]
\end{equation}
the first terms of which one recognises as the continuum (lattice)
functionals whose stationary points with
respect to gauge transformations give the Landau gauge.  

The detailed form of the continuum Lagrangian after implementing
the algebra has been given 
elsewhere \cite{Curci:1976,Thierry-Mieg:1985}.
We merely highlight specific terms.
Firstly, after integration by parts,
${\bar s} \int d^4x A^2 = - 2 \int d^4x {\bar C} \partial_{\mu} A_{\mu}$.
Acting with $s$ then gives the ghost kinetic term
and multiplier field term $B \partial_{\mu} A_{\mu}$,
which are standard in covariant gauges. 
Secondly, the most complicated structures emerge from
$s{\bar s} ({\bar C} C)$. The damping term
for the multiplier field $B^2$ emerges from here,
also as in standard covariant gauges. But 
one ghost mass term, three-point
couplings as well as quartic couplings $({\bar C} C)^2$ and permutations thereof appear.
Finally the $m^2 (A^2-\xi {\bar C} C)$ generates both
a gluon and ghost mass term.  

Let us consider now some typical expectation
value of a BRST invariant observable
of the link fields $O[U]$ in the lattice theory corresponding to the
massive Curci-Ferrari gauge:
\begin{equation}
\langle O[U]\rangle_{mCF}= 
{{\int DU D\phi e^{-S_{YM}[U]-S_{GF}[U,\phi]}O[U]}\over
{{\int DU D\phi e^{-S_{YM}[U]-S_{GF}[U,\phi]} }}}
\end{equation}
where $\phi$ represents the auxiliary fields, $C, \bar{C}$ and $B$.
We shall now examine the precise relationship between
this expectation value and that for lattice YM theory
which, in terms of link variables, is well-defined
even without gauge-fixing. We can factor into
numerator and denominator the finite (on the lattice)
integration over the gauge group, $V_G=\int Dg < \infty$
and use the standard trick of exploiting the invariance
of the measure, observable and YM action under gauge transformations to
rewrite the expectation value as
\begin{eqnarray}
\langle O[U]\rangle_{mCF} =
 {{\int DU e^{-S_{YM}[U]}O[U] {\bar Z}[U]}\over
{\int DU e^{-S_{YM}[U]} {\bar Z}[U]}}
\end{eqnarray}
where
\begin{equation}
{\bar Z}[U]= \int Dg D\phi e^{-S_{GF}[U^g,\phi]}
\label{GFQFT}
\end{equation}
represents the partition function of a field theory
in the group $g$ and auxiliary variables $\phi$
in the background of the link field $U$. 
{\it But this is not a topological quantum field theory}
because the action of this theory is not based on a nilpotent
algebra \cite{Birmingham:1991}. 
The consequence of this is that, unlike for TQFTs,
this partition function depends on the background field,
${{\delta {\bar Z}[U]}\over {\delta U}} \neq 0$.
The proof can be sketched as follows.
Since the measure of ${\bar Z}[U]$ is independent of the
link field the variation with respect to $U$ acts directly
onto the exponential of the action of the theory, $S_{GF}[U^g,\phi]$,
bringing the action into the measure. The variation $\delta/\delta U$
commutes with the operation $s{\bar s}-m^2$ so that we effectively have
\begin{equation}
\int Dg D\phi (s{\bar s}-m^2) {{\delta W}\over {\delta U}}
e^{-S_{GF}[U^g,\phi]}.
\end{equation}
But the integral of a BRST exact quantity with respect
to an invariant measure still vanishes, despite the
lack of nilpotency, thus
\begin{equation}
{{\delta {\bar Z}[U]}\over {\delta U}} = m^2 \int Dg D\phi
{{\delta W}\over {\delta U}}
e^{-S_{GF}[U^g,\phi]}
\end{equation}
which is not evidently constrained to vanish by any symmetry argument.
Thus the partition function depends on the
background link. Thus ${\bar Z}[U]$ cannot be factored out
and cancelled except in the massless limit, so that
YM expectation values can only be defined via the limit 
\begin{equation}
\langle O[U] \rangle_{YM} = \lim_{m\rightarrow 0} \langle O[U]\rangle_{mCF}.
\end{equation}
Unfortunately, the usual tricks of TQFT cannot help us in
explicitly evaluating ${\bar Z}[U]$ here: 
because of explicit dependence on $U$ neither a semiclassical limit
can be taken (as in TQFT) nor can the trivial link $U=1$
be chosen (which for the lattice theory would result in some
spin model \cite{Schaden:1998}).
We argue however that for {\it generic} $m\neq 0$ the partition function 
${\bar Z}[U]$ will be non-vanishing:
the functional being introduced into the
measure {\it a la} Faddeev-Popov trick is {\it orbit-dependent}
and this is precisely what we require in order to
lift the degeneracy between Gribov regions to avoid the
Neuberger problem. We shall illustrate this below
for the case of a simple one-link model.

At any rate, we can give a final formula for the expectation
value of a gauge-invariant observable in Yang-Mills theory
in terms of the present construction:
\begin{equation}
\langle O[U] \rangle_{YM} = \lim_{m\rightarrow 0}
{{\int DU e^{-S_{YM}[U]}O[U] {\bar Z}[U]}\over
{\int DU e^{-S_{YM}[U]} {\bar Z}[U]}}
\label{finalformula}
\end{equation}
with ${\bar Z}[U]$ defined by Eq.(\ref{GFQFT}). 
We can use the language of soft-meson theorems where the
pion is a pseudo-Goldstone boson for massive quarks and  
thus describe ${\bar Z}[U]$ as the partition function of 
a {\it pseudo}-topological quantum field theory PTQFT.

We can now elucidate how the Neuberger problem is avoided.
Neuberger \cite{Neuberger:1986} considers the integral
\begin{equation}
I_{O}(t)= \int D\phi e^{-S_{0} -t s F} O[U,\phi]. 
\end{equation}
The measure $D\phi$ is BRST invariant as is the action $S_{0}$,
which includes the Yang-Mills action.  
Expectation values in the theory are obtained for  
$t=1$, namely $\langle O \rangle = I_{O}(1)/I_{1}(1)$.
However $S_0$ must also contain damping terms for the 
scalar $B$ field integrations
in $D\phi$ since even on the lattice these field directions
(unlike the link field $U$) are not compact. 
The damping term must be itself BRST invariant
(upon an appropriate shift of the $B$-field --
actually in the standard case it is even BRST exact,
$B^2=s({\rm{something}})$ but this is not so relevant here).
Variation of $I_{O}(t)$ with respect to $t$
brings $sF$ into the measure, the integrals of which vanish.
Thus $dI_{O}/dt=0$ and $I_O(t)$ is t-independent. 
But for $t=0$ one has an integrand containing
no ghost fields which vanishes by the rules of Grassmann integration:
$\int DC=0$. Thus $I_O(0)=0=I_O(1)$ and all expectation values
are of the form $0/0$. {\it This assumes $I$ is well-defined
at each step, which is only the case if $S_0$ contains the
damping term for $B$}. For the massive-Curci-Ferrari case
the structure of the action is different
on two grounds: $B^2$ is not BRST exact,  
$B^2 \neq s({\rm {something}})$,
but more importantly
shifting $B$ to $b=B+\frac{1}{2} \{ {\bar C}, C \}$ gives
\begin{equation}
s {\rm{Tr}}\ b^2 = 2 m^2 {\rm{Tr}}( C b),
\end{equation}
so that the damping term cannot be placed in $S_0$
but must be placed in the term multiplied by $t$. There are two ways to
do this, but keeping as close as possible to Neuberger's
original argument we can reassign $s{\bar s}W \rightarrow t s{\bar s}W $
such that under derivation with respect to $t$ a
BRST exact term comes down into the measure. However, either
way we cannot consider the $t=0$ limit as the functional
integral then becomes undamped. This makes the Neuberger limit $t\rightarrow 0$
fail, and so the usual proof fails.

We now explicitly study this proposal in
the context of the simple model introduced by Testa \cite{Testa:1998}.
We consider an Abelian model with only two lattice sites,
$x_1,x_2$ and thus only one degree of freedom,
a link variable $U$ which is parametrised through its phase
$U=e^{i a A}$. $A$ is compact, $A \in [-\frac{\pi}{a},\frac{\pi}{a})$.
A gauge transformation corresponds to shifting $A$ by a
difference $(\omega(x_2)-\omega(x_1))/a$ which is a fixed
quantity for any function $\omega(x)$. There are no plaquettes
so the action is zero. The model essentially only contains
topological information.
The BRST and anti-BRST algebras for this simple gauge field theory can 
be written
\begin{eqnarray}
s A &=& C , \ {\bar s} A = {\bar C} \nonumber \\
s C &=& 0 , \ {\bar s} {\bar C} = 0 \nonumber \\
s {\bar C} &=& i(B + {\bar C} C) , \ {\bar s} C =i( - B + {\bar C} C )
\nonumber \\
s B &= & -i (m^2+B)C, \ {\bar s} B = -i (m^2-B){\bar C}.
\end{eqnarray}
Note that $B$ is now a real field, 
ghosts are Maurer-Cartan one-forms and
\begin{eqnarray}
s^2 {\bar C} = m^2 C, \
{\bar s}^2 C = - m^2 {\bar C},
\end{eqnarray} 
so nilpotency is lost.
Using that $s V[A] = V^{\prime} C$ and ${\bar s} V[A] = {\bar C} V^{\prime}$
the gauge-fixing action for the massive Curci-Ferrari model here
gives
\begin{eqnarray}
{S}_{gf} &=& (s{\bar s} - m^2) \left( V[A] - \xi {\bar C} C \right)
\nonumber \\
&=& {\bar C} [ - V^{\prime\prime} + iV^{\prime} + 2 \xi B +2 m^2 \xi] C
+\xi B^2 + i B V^{\prime} - m^2 V
\end{eqnarray}
where $V[A]$ is constrained only by the requirement of periodicity
under $A\rightarrow A+2\pi/a$. Now this action
appears Gaussian in all fields because quartic terms ${\bar C}^2 C^2$
all vanish since there is only one species of Grassmann field.
Integration out of either ghosts or scalar field $B$ will upset this.
In this case we can shift $B$ via $b=B+{\bar C} C$ and avoid this.
The action then becomes
\begin{equation}
{\bar C}[-V''+2m^2 \xi] C + i b V' + \xi b^2 - m^2 V
\end{equation}
and integration out of $b$ will give the Gaussian  
$\sqrt{\pi}\exp(-\frac{1}{4\xi}(V'[A])^2)/\sqrt{\xi}$.
The procedure is dependent on the gauge parameter $\xi$ for the same
reasons as ${\bar Z}$ is $U$-dependent, but we will consider
the case closest to the Landau gauge, 
$\xi\rightarrow 0$ for which we obtain then the delta function
$\delta[V']$ in the measure. Integrating out
ghost fields gives for the partition function of the
original theory:
\begin{equation}
Z = \int_{-\pi/a}^{\pi/a}dA (-V''[A]) \delta[V'[A]] e^{m^2 V[A]}
\end{equation}
which should be compared with the partition function
for the unfixed theory: $\int_{-\pi/a}^{\pi/a} dA=2\pi/a$.

We see that only stationary points of $V[A]$ contribute.
They are weighted by the second derivative
which would otherwise correspond to the Faddeev-Popov determinant; 
both signs of $V''$ can appear. We crucially
see the additional orbit-dependent weighting exponential in the mass  
consistent with our
observations above. If there are many such stationary points -
Gribov copies - all of them will be summed over. But since
$V[A]$ is not a gauge-invariant functional of $A$, they will
generally come with different weight unless they represent
degenerate stationary points of $V[A]$. 
For $m=0$ we recover the Neuberger pathology:
critical points cancel according to the sign
of the second derivative. The weight factor $e^{m^2 V[A]}$ breaks this
degeneracy so that the partition function will not vanish
and the $0/0$ problem disappears.

Let us see this more explicitly.
For simplicity we work in units of lattice spacing now $(a=1)$
and choose $V[A]=\frac{1}{2} \sin^2 A $ 
so that $V'[A]=\frac{1}{2} \sin 2A$ and $V''[A]=\cos 2A$.
The ``gauge-fixed'' configurations are thus
\begin{equation}
A_{{\rm{fixed}}} = -\pi, -\frac{\pi}{2}, 0, \frac{\pi}{2}
\end{equation}
for which $V''$ has values $1,-1,1,-1$ respectively
while the original $V$ has values $0,\frac{1}{2},0,\frac{1}{2}$. 
The partition function is trivially evaluated to be
\begin{eqnarray}
Z = 2 (e^{\frac{m^2}{2}}-1) \approx m^2.
\end{eqnarray}
Note that the constraint that observables $O$ be BRST invariant 
in this case means that $dO[A]/dA=0$, thus BRST invariant observables
are just constants, $O=c$.
All integrals of $O$ are just multiples of the partition function, $c Z$.
Thus the expectation value $\langle O[A] \rangle=c$ in the unfixed theory.
Setting $m=0$ before taking the ratio reintroduces 
the pathological $0/0$ result
for the expectation value. Conversely, keeping $m$ small but finite 
while taking the ratio for the expectation value allows 
$m^2$ factors to safely cancel in numerator and denominator,
giving the correct result for the expectation value, namely the
constant $c$.

Whether such an elegant cancellation in expectation values
of observables can take place
in general, or subtleties of the $m\rightarrow0$ limit
need to be taken into consideration is an open question.
For example, the vanishing of 
the partition function at $m=0$ can also be understood as
the vanishing of the Witten index of the underlying  
TQFT with its supersymmetry. This leaves
open the possibility of spontaneous breaking of BRST
symmetry which would jeopardise the BRST cohomology construction of
a physical state space.  
This aside, there is the question
whether renormalisation effects can hinder the program.
This is related to the other limits which must follow a lattice calculation:
the continuum $(a\rightarrow 0)$  and thermodynamic $(L\rightarrow \infty)$
limits. Note that for a finite lattice, the lattice gauge group
is a simple product of $SU(n)$ gauge groups per lattice site.
Thus the original Neuberger zero corresponds to $0^{\rm{no. \ of \ sites}}$
which goes over to a single zero in the $a\rightarrow 0$ limit,
related to the remaining global gauge symmetry of the torus at finite volume
\cite{Schaden:1998}. In order to avoid the Neuberger problem
reappearing in our proposal it is essential that the Curci-Ferrari mass
remain non-zero for finite $L$ but vanishing $a$,
particularly in light of renormalisation of the mass. For sufficiently
large lattice volume, there will be no $L$ dependence
in the appropriate (and only UV sensitive) renormalisation constants for $m$;
for example \cite{deBoer:1996,Gracey:2001} have computed these in 
continuum perturbation theory. Thus renormalisation will not
introduce $L$-dependence into $m$. This suggests that it
is safe to take $m\rightarrow 0$ either as fast as $L\rightarrow\infty$
or independently. But we stress that these considerations are only heuristic.

We reiterate that unlike approaches seeking to 
isolate the Gribov or fundamental modular region
in the space of gauge fields, this approach takes all Gribov regions
into account. 
There is the proposal  
that in the {\it infinite volume limit} configurations on the common
boundary of Gribov and fundamental modular regions
will dominate the ensemble averages of gauge
invariant observables \cite{Zwanziger:2003}. 
However such ``dominant'' fields will always be
some subset of those configurations 
contributing at finite $(L,a)$ thus representing
no contradiction. This is also explicitly evident in the simple model.

Open work includes careful examination of the $m\rightarrow 0$ limit
for the full theory. We mention here that the
violation of nilpotency of the BRST algebra results in
negative norm states appearing in the physical Hilbert space
\cite{Ojima:1982,deBoer:1996} defined according to
the Kugo-Ojima criterion \cite{Nakanishi:1990}. This is one way of
manifesting the loss of unitarity. 
These states do not belong to Kugo-Ojima quartets. In \cite{Ojima:1982} 
one sees explicitly that in the massless limit they become zero norm 
(``daughter'') states.
Thus from the point of view of the Hilbert space, the massless
limit is smooth, as it is also for perturbative
Green's functions. However, whether the same can be
said for non-perturbative Green's functions
is an open challenging question.
There also remain technical challenges to
implementing quartic ghost couplings in the lattice
framework. The introduction of additional auxiliary bosonic
fields, as in Nambu--Jona-Lasinio models, may be a way forward
in this problem. 

To conclude, we have shown that the
massive Curci-Ferrari model overcomes the Neuberger problem
for elevating BRST symmetry to the non-perturbative level on
a finite lattice. Beyond this first step,
the verification that the massless,  
continuum, and thermodynamic limit of this procedure is the 
physically relevant theory faces a number of difficult challenges still.

\section*{Acknowledgements}
ACK and LvS wish to acknowledge the support
of both the CSSM and the ITPIII for mutual visits.
ACK is supported by the Australian Research Council.
This work was supported by the DFG under grant SM 70/1-1.
We are indebted to 
discussions with Marco Ghiotti, Peter Bouwknegt,
Mark Stanford, Massimo Testa,
Peter Jarvis, Bob Delbourgo and Jan-Willem van Holten.




\begin{thebibliography}{00}
\bibitem{Becchi:1975}
C.~Becchi, A.~Rouet and R.~Stora,
Annals Phys.\  {\bf 98} (1976) 287;
I.~V.~Tyutin,
LEBEDEV-75-39
\bibitem{Alkofer:2000}
R.~Alkofer and L.~von Smekal,
Phys.\ Rept.\  {\bf 353} (2001) 281
[arXiv:hep-ph/0007355].
\bibitem{Bowman:2004}
P.~O.~Bowman, U.~M.~Heller, D.~B.~Leinweber, M.~B.~Parappilly 
and A.~G.~Williams,
Phys.\ Rev.\ D {\bf 70} (2004) 034509
[arXiv:hep-lat/0402032].
\bibitem{Neuberger:1986}
H.~Neuberger,
Phys.\ Lett.\ B {\bf 183} (1987) 337.
\bibitem{Gribov:1977}
V.~N.~Gribov,
Nucl.\ Phys.\ B {\bf 139} (1978) 1.
\bibitem{Dell'Antonio:1991}
G.~Dell'Antonio and D.~Zwanziger,
Commun.\ Math.\ Phys.\  {\bf 138} (1991) 291.
\bibitem{Baulieu:1996}
L.~Baulieu and M.~Schaden,
Int.\ J.\ Mod.\ Phys.\ A {\bf 13} (1998) 985
[arXiv:hep-th/9601039].
\bibitem{Schaden:1998}
M.~Schaden,
Phys.\ Rev.\ D {\bf 59} (1999) 014508
[arXiv:hep-lat/9805020].
\bibitem{Golterman:2004}
M.~Golterman and Y.~Shamir,
arXiv:hep-lat/0404011.
\bibitem{Curci:1976}
G.~Curci and R.~Ferrari,
Nuovo Cim.\ A {\bf 35} (1976) 1
[Erratum-ibid.\ A {\bf 47} (1978) 555].
\bibitem{Baulieu:1981}
L.~Baulieu and J.~Thierry-Mieg,
Nucl.\ Phys.\ B {\bf 197} (1982) 477.
\bibitem{Thierry-Mieg:1985}
J.~Thierry-Mieg,
Nucl.\ Phys.\ B {\bf 261} (1985) 55.
\bibitem{Alkofer03}
R.~Alkofer, C.~S.~Fischer, H.~Reinhardt and L.~von Smekal,
Phys.\ Rev.\ D {\bf 68} (2003) 045003
[arXiv:hep-th/0304134].
\bibitem{Birmingham:1991}
D.~Birmingham, M.~Blau, M.~Rakowski and G.~Thompson,
Phys.\ Rept.\  {\bf 209} (1991) 129.
\bibitem{Testa:1998}
M.~Testa,
Phys.\ Lett.\ B {\bf 429} (1998) 349
[arXiv:hep-lat/9803025].
\bibitem{deBoer:1996}
J.~de Boer, K.~Skenderis, P.~van Nieuwenhuizen and A.~Waldron,
Phys.\ Lett.\ B {\bf 367} (1996) 175
[arXiv:hep-th/9510167].
\bibitem{Gracey:2001}
J.~A.~Gracey,
Phys.\ Lett.\ B {\bf 525} (2002) 89
[arXiv:hep-th/0112014].
\bibitem{Zwanziger:2003}
D.~Zwanziger,
Phys.\ Rev.\ D {\bf 69} (2004) 016002
[arXiv:hep-ph/0303028].
\bibitem{Ojima:1982}
I.~Ojima,
Z.\ Phys.\ C {\bf 13} (1982) 173.
\bibitem{Nakanishi:1990}
N.~Nakanishi and I.~Ojima,
``Covariant Operator Formalism Of Gauge Theories And Quantum Gravity,''
World Sci.\ Lect.\ Notes Phys.\  {\bf 27} (1990) 1.

\end{thebibliography}
\end{document}